\documentclass{elsart}

\usepackage{epsfig}
\begin{document}

\begin{frontmatter}



\title{Anomalous stress relaxation in random macromolecular networks}


\author{Kurt Broderix\corauthref{Kurt},}
\author{Henning L\"owe, Peter M\"uller and Annette Zippelius}

\corauth[Kurt]{Deceased 12 May 2000}

\address{Institut f\"ur Theoretische Physik,
  Georg-August-Universit\"at, D--37073 G\"ottingen, Germany}

\begin{abstract}
Within the framework of a simple Rouse-type model we present exact
analytical results for dynamical critical behaviour on the sol side of
the gelation transition. The stress-relaxation function is shown to
exhibit a stretched-exponential long-time decay. The divergence of the
static shear viscosity is governed by the critical exponent $k=\phi
-\beta$, where $\phi$ is the (first) crossover exponent of random
resistor networks, and $\beta$ is the critical exponent for the gel
fraction. We also derive new results on the behaviour of normal stress
coefficients. 
\end{abstract}



\end{frontmatter}

\section{Introduction}
\begin{figure}
  \begin{minipage}[t]{0.47\textwidth}
    \begin{center}
      \epsfig{file=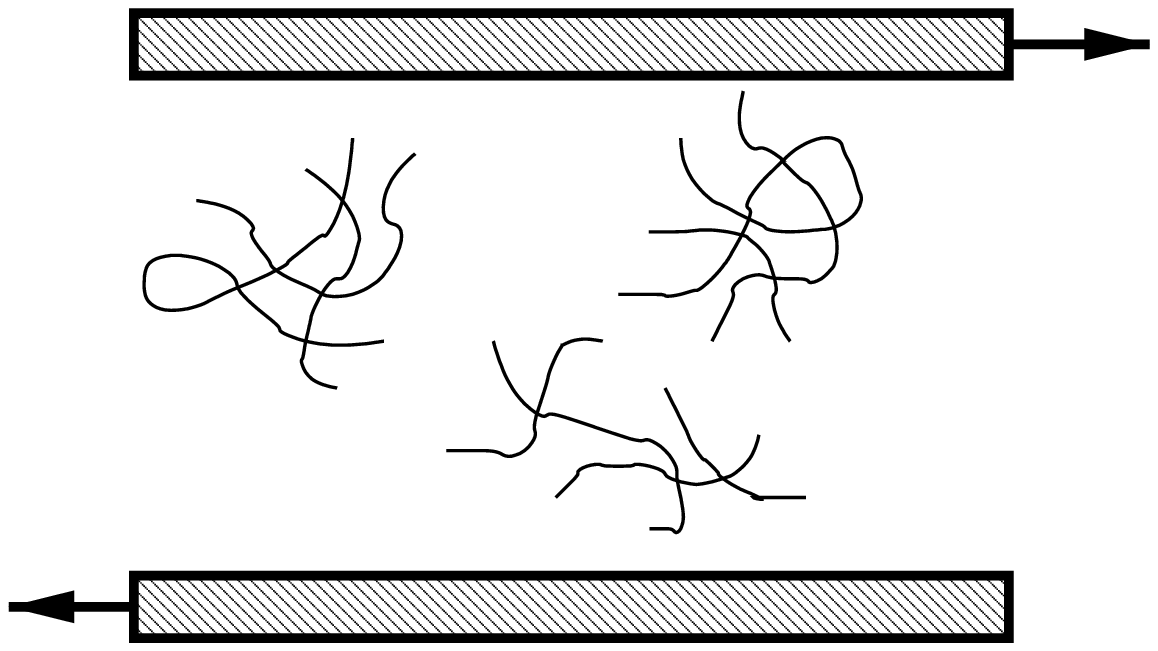, width=\textwidth}
    \end{center}
    \caption{Polymers in a shear experiment}
    \label{fig:compareA}
  \end{minipage}\hfill
  \begin{minipage}[t]{0.47\textwidth}
    \begin{center}
      \epsfig{file=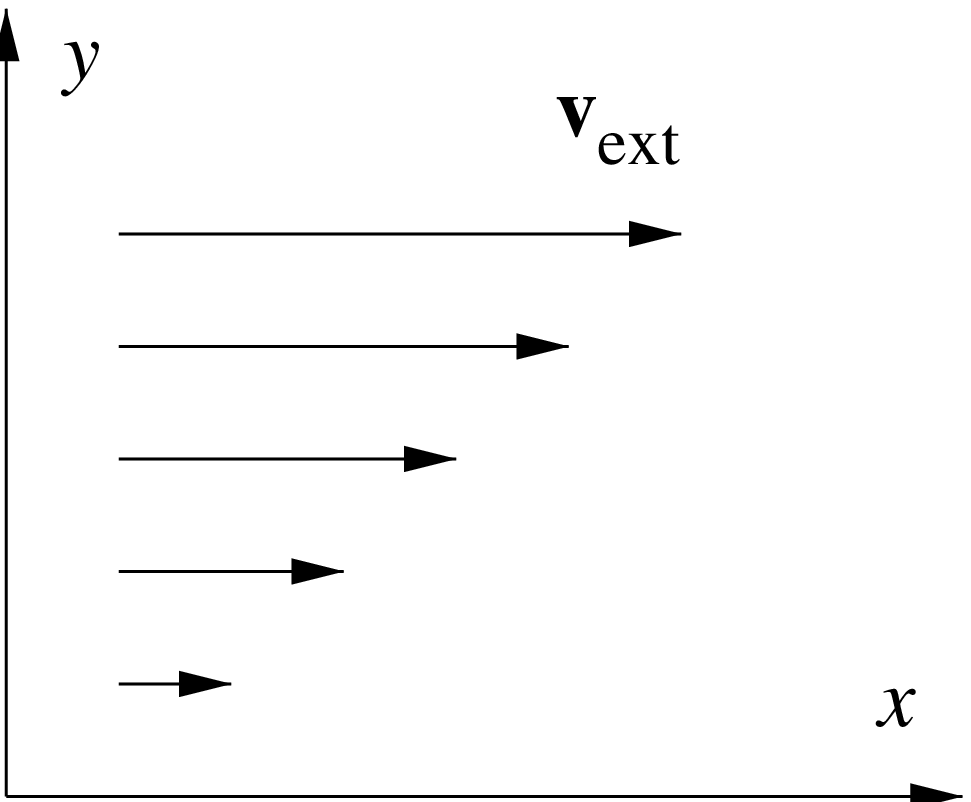, width=0.7\textwidth}
    \end{center}
    \caption{Homogeneous linear shear flow (\protect{\ref{flowfield}})}
    \label{shearflow}
  \end{minipage}
\end{figure}

The viscoelastic properties of incipient gels have received
considerable interest, but are still controversial. Whereas the static
critical behaviour as predicted by percolation theory has been
confirmed in experiment as well as simulation \cite{StCo82}, the
dynamic critical 
behaviour, and in particular stress relaxation is much less
understood.  Conceptually, the experimental procedure is depicted in
Figs.~\ref{fig:compareA} and~\ref{shearflow}: a
homogeneous, time-dependent shear flow is imposed, and the
stress relaxation is measured. Throughout the sol phase,
even far away from the critical point, one observes an anomalous time
decay of the shear-stress relaxation function, which follows a
stretched-exponential law $\exp\{-(t/t^{*})^{\alpha}\}$. 
However the exponent values
\cite{AdDe85,DuDe87,Win87,MaAd88,MaAd89,AdMa90} vary over a wide
range, and in some  
experiments even non-universal exponents are reported, 
depending for example on molecular weight or concentration
of crosslinks.  Similarly, the divergence of the static shear
viscosity is still a matter of debate. Experiments yield wide ranges
of exponent values \cite{AdDe81,MaAd88,CoGi93} whose origin is not
clear. Even the existence of a single dynamic universality class has
been questioned \cite{AdDe81}. 

From a theoretical point of view, the issue is not clear either.
Scaling arguments which are 
based on simple physical pictures or analogies to other random
systems, are in conflict with each other.  For example, it has been
argued \cite{Gen78} that the static shear viscosity $\eta_n$ of a
cluster of $n$ polymers should be determined by the longest relaxation
time $t_n$ of 
the cluster. The latter has been identified with the time scale for
diffusion of the cluster over a distance corresponding to its own
size: $t_n \propto R_n^2/D_n$. Here $R_n$ denotes the radius of gyration
that is assumed to scale with the mass of the cluster as $R_n \propto
n^{1/d_{f}}$ with $d_{f}$ the Hausdorff-Besicovich dimension.  In a simple
Rouse model the diffusion constant behaves as $D_n \propto 1/n$. The
probability 
to find a cluster of size $n$ is taken from percolation theory to
yield $\eta\propto \varepsilon^{-k}$ with $k= 2\nu - \beta$. Here
$\varepsilon$ denotes the distance from the critical point and $\nu$ and
$\beta$ are the exponents of percolation theory for the correlation
length and the gel fraction, respectively. In another line of approach
one has tried 
to relate the viscosity to random resistor networks. For example, it
was suggested \cite{Gen79} that the conductivity of a random mixture of
conductors and superconductors should show the same critical behaviour
as the viscosity, $s=k$.

In this Paper we discuss Rouse dynamics in an externally imposed shear
flow, generalised to include the effects of random, permanent
crosslinks. Within this model we have recently found \cite{BrLo99,BrLo01} the 
exact result $k=\phi-\beta$ for the critical divergence of the 
static shear viscosity.
Here $\phi$ is the (first) crossover exponent for random resistor
networks, which is discussed e.g.\ in \cite{HaLu87,StJaOe99}. This
analytical result is in contradiction with 
all previous scaling arguments, but agrees with that of a
molecular-dynamics simulations \cite{VePl01}. The afore-mentioned 
scaling arguments fail, because they ignore the multi-fractal structure
of percolation clusters -- as first noted by Cates \cite{Cat85}. 
To account for multi-fractality one needs to introduce another fractal
dimension, the spectral dimension $d_s$, which  is independent of
$\beta$ and $\nu$. The crossover exponent $\phi$ is 
related to $d_{s}$ according to $\phi=\nu d_{f}
[(2/d_s)-1]$ so that our result $k=\phi-\beta$ is incompatible with
the proposal $k= 2\nu - \beta$. It is also incompatible with the other
suggestion $s=k$, as can be seen most easily in $d=2$, where duality 
implies $s=\phi$ \cite{Han90}.

Whereas for Newtonian fluids the simple shear flow of
Fig.~\ref{shearflow} gives rise only to a shear stress $\sigma_{xy}$,
is is well known that in a non-Newtonian fluid all six components of
the stress tensor are non-zero \cite{BiAr87}. If the fluid is
incompressible and isotropic, there are three independent,
experimentally observable stress components: the shear stress
$\sigma_{xy}$, the first normal stress difference
$\sigma_{xx}-\sigma_{yy}$ and the second normal stress difference
$\sigma_{yy} -\sigma_{zz}$. We compute the normal stress coefficients
$\Psi_{1}$ and $\Psi_{2}$ and find $\Psi_{2}=0$, a characteristic result for
Rouse-type models. The first normal stress coefficient $\Psi_{1}$ is
predicted to have a much stronger 
divergence than the shear viscosity as the gelation transition is
approached. Even though one has performed many
experiments \cite{LoMe73} on both the shear-rate dependence 
of normal stresses in entangled or (temporarily) crosslinked polymeric liquids
and on the time dependence of the normal-stress response to
particular shapes of shear strain, we are not aware of any experiments
measuring $\Psi_{1}$ as a function of the crosslink concentration.
Previous theoretical work \cite{WiMo97} relates the critical
divergence of $\Psi_{1}$ at the gelation transition to scaling
properties of the
relaxation-time spectrum, see also \cite{BrMu01} for a
recent approach in a similar spirit.

Finally, we present a result on the long-time decay of the
shear-stress relaxation function. We show that this decay is described
by a stretched exponential, which is determined by the soft-mode
excitations of the clusters, see \cite{BrAs00} for details.

\section{Model and Observables}

We consider a fluid of $N$ identical molecular units each consisting
of $L$ monomers. Examples are chains or rings of length $L$ or stars
with $(L-1)$ branches. Within the Rouse-type model studied in this
Paper it turns out that the critical behaviour as
well as the anomalous long-time decay in the sol phase is independent
of the internal structure of the molecular units. We therefore only
discuss the simplest case in detail, namely molecular units which are
just monomers, corresponding to $L=1$. The reader who is interested in
results for more complex molecular units is referred to
\cite{BrLo99,BrLo01}. 

\begin{figure}
  \begin{minipage}[c]{0.39\textwidth}
    \begin{center}
      \epsfig{file=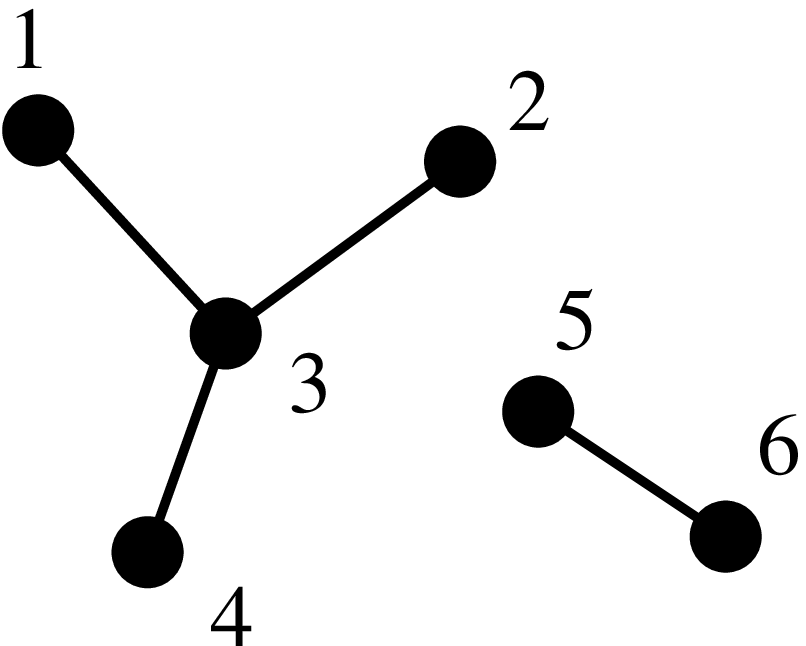, width=0.7\textwidth}
    \end{center}
  \end{minipage}
  \begin{minipage}[c]{0.59\textwidth}
    \begin{center}
      \epsfig{file=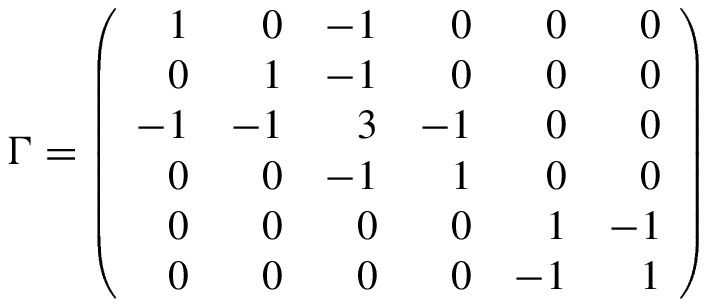, width=\textwidth}
    \end{center}
  \end{minipage}
  \caption{A particular crosslink realisation and the associated
    connectivity matrix \protect{$\Gamma$} (with all coupling
    constants \protect{$\lambda_{e}=1$}).}
  \label{gammaex}
\end{figure}

The thermal degrees of freedom are the positions $\mathbf{R}_i(t)$,
$i=1,\ldots,N$, of the monomers, which relax in the presence of $M$
quenched, random crosslinks, each connecting a pair $(i_e,i'_e)$,
$e=1,\ldots,M$, of monomers. Crosslinks are modelled as harmonic springs
\begin{equation} \label{Eq4}
  U := \frac{3}{2a^2}\:\sum_{e=1}^M \lambda_e
  \bigl( \mathbf{R}_{i_e}-\mathbf{R}_{i'_e} \bigr)^2,
\end{equation}
with random spring constants $\lambda_{e}$ and an overall coupling
strength determined by the length $a>0$. It is advantageous to express
the potential energy in terms of the connectivity matrix 
\begin{equation}
   \label{Eq4_1}
      \Gamma_{ii'}    : = \sum_{e=1}^{M} \lambda_e
    (\delta_{ii_{e}} - \delta_{ii_{e}'})
    (\delta_{i'i_{e}}- \delta_{i'i_{e}'})
\end{equation}  
according to
\begin{equation}
U= \frac{3}{2a^2}\:\sum_{i,i'=1}^{N} \Gamma_{ii'}\,  \mathbf{R}_i\cdot
\mathbf{R}_{i'} \,.
\end{equation}
A specific realisation of crosslinks is represented as a graph $\mathcal{
  G}=\{i_e,i_{e}'\}_{e=1}^{M}$ or, equivalently, by its connectivity matrix
$\Gamma$. A simple example is shown in Fig.~\ref{gammaex}.

We consider Rouse dynamics generalised to include the effects of
permanent random crosslinks \cite{DoEd85,BiCu87,SoVi95} and an
externally applied 
velocity field ${\mathbf{v}}_{\mathrm{ext}}({\mathbf{r}},t)$ 
\begin{equation} \label{eqmotion}
 \zeta \left[ \partial_t R^{\alpha}_i(t)
- v^{\alpha}_{\mathrm{ext}}\bigl(\mathbf{R}_i(t),t\bigr)\right] = 
  - \frac{\partial U}{\partial R^{\alpha}_i}(t) 
  + \xi^{\alpha}_i(t).
\end{equation}
Here, Greek indices label Cartesian coordinates $x$,$y$ or $z$.
Inertial terms are neglected in (\ref{eqmotion}), and friction with a
friction constant 
$\zeta$ occurs when the velocity of a monomer deviates from the
externally applied flow field. The crosslinks exert a force $-\partial
U/\partial\mathbf{R}_i$ on the monomers, in addition to a random,
fluctuating thermal-noise force with zero mean and covariance
$\langle\xi_{i}^{\alpha}(t)\xi_{j}^{\beta}(t')\rangle = 2\zeta
\delta_{\alpha\beta}\delta_{ij}\delta(t-t')$. Note that we have chosen
units in which the inverse temperature is equal to one.

In the sequel we will only be interested in a homogeneous linear shear
flow 
\begin{equation}
  \label{flowfield}
  v^{\alpha}({\mathbf{r}},t) := \delta_{\alpha x}\,\kappa(t)\, y
\end{equation}
with a time-dependent shear rate $\kappa(t)$, which is sketched in
Fig.~\ref{shearflow}.  
Given this shear flow, the equation of motion (\ref{eqmotion}) is
linear and can be solved exactly for each realisation of the thermal
noise \cite{BrLo01}.

In reaction to the externally applied shear flow, the crosslinked
polymer system exhibits stress, whose tensor components are given in
terms of a force-position correlation \cite{DoEd85,BiCu87}
\begin{equation}
  \sigma_{\alpha\beta}(t) = \lim_{t_{0}\to -\infty}
  \frac{\rho_{0}}{V}\,\sum_{i=1}^{N} \left\langle\frac{\partial U}{\partial
    R_{i}^{\alpha}}(t) \; R_{i}^{\beta}(t)\right\rangle\,.
\end{equation}
Here, $\rho_{0}$ denotes the density of monomers, and the initial values
are to be taken at more and more distant times $t_{0}$
in the past in order to ensure that after averaging over the thermal
noise, the system has reached a steady state at time $t$. This yields
for the stress tensor
\begin{equation}
\label{stresstensor}
\mbox{\boldmath$\sigma$} (t) = \chi(0)\mathbf{1} + \int_{-\infty}^{t}\!\d t' \,
\chi(t-t')\, \kappa(t') \left(\begin{array}{cc@{\qquad}c}
2\int_{t'}^{t}\!\d s\,\kappa(s) & 1 & 0 \\
1 & 0 & 0\\ 0 & 0 & 0\end{array}\right)\,,
\end{equation}
confer \cite{BrLo01}, where the stress relaxation function is given by
\begin{equation}
\label{eq:stressrelax}
\chi(t)=
\frac{\rho_{0}}{N}\;\mathrm{Tr} \left([1-E_0(\mathcal{ G})] 
\exp\left\{-\frac{6t}{\zeta a^2}\Gamma(\mathcal{ G})\right\}\right)\,,
\end{equation}
and the dependence of $\Gamma$ on the crosslink realisation
$\mathcal{G}$ has been emphasised in the notation. In (\ref{eq:stressrelax}),
$E_0$ denotes the projector on the space of zero
eigenvalues of $\Gamma$. These zero eigenvalues correspond to
translations of whole clusters. The associated eigenvectors are
constant within one cluster and zero outside. In
the example of Fig.~\ref{gammaex} the null space is spanned by the
vectors $\mathbf{a}_{1}:=4^{-1/2}(1,1,1,1,0,0)$ and
$\mathbf{a}_{2}:=2^{-1/2}(0,0,0,0,1,1)$, and one has $E_{0}=
\mathbf{a}_{1}\mathbf{a}_{1}^{\mathsf{t}} +
\mathbf{a}_{2}\mathbf{a}_{2}^{\mathsf{t}}$.  
Within the simple Rouse model the zero eigenvalues do not
contribute to the shear relaxation because there is no force acting
between different clusters. The only contribution to stress relaxation
is due to deformations of the clusters, as can be seen from
(\ref{eq:stressrelax}). The long-time decay of $\chi$ will be
dominated by the smallest eigenvalues of $\Gamma$, and we 
expect that anomalies in the long-time behaviour are due to
a peculiar behaviour of the density of eigenvalues of $\Gamma$ for
small eigenvalues. 

To complete the definition of the dynamic model, we need to specify
the distribution of random crosslinks. Two cases will be distinguished: 
\begin{enumerate}
\item[(i)] Each pair of monomers is chosen with equal probability
  $c/N$, generating mean-field random graphs as discussed by Erd{\H o}s and
  R\'enyi \cite{ErRe60}, see also \cite{Bol98}. As a 
  function of crosslink concentration $c$ , the system undergoes a
  percolation transition at a critical concentration
  $c_{\mathrm{crit}}= \frac12$.
  For $c<c_{\mathrm{crit}}$ there is no macroscopic cluster, and almost all
  clusters are trees. The average number of tree clusters per particle
  is given in the macroscopic limit by
  \begin{equation}\label{Eq10_3}
    \tau_{n}=
    \frac{n^{n-2}(2c\,\e^{-2c})^n}{2c\,n!}. 
  \end{equation}
\item[(ii)] Crosslinks are distributed such that the cluster-size
  distribution follows a scaling law 
  \begin{equation} 
    \label{scaling}
    \tau_n=n^{-\tau}f\bigl((c_{\mathrm{crit}}-c)n^{\sigma}\bigr)
  \end{equation}
  near criticality. This case includes both mean-field random graphs and
  random bond percolation. 
\end{enumerate}

For both cases (i) and (ii) we require the random spring
constants $\lambda_{e}$ to be distributed independently from each
other, as well as independently from the crosslink positions. Moreover, the
probability for  
very soft spring constants to occur shall be sufficiently small in
that sufficiently high inverse moments 
\begin{equation}
  \label{moments}
  P_{n}:= \int_{0}^{\infty}\!\d\lambda\, \lambda^{-n} p(\lambda)
\end{equation}
of $\lambda_{e}$ are assumed to exist. 

The combined average over crosslink configurations and random spring
constants will be denoted by an overbar
$\overline{\phantom{l}{\scriptstyle\bullet}\phantom{l}}$. Using
this notation, we implicitly assume that the macroscopic limit 
$N\to\infty$, $M\to\infty$, $M/N\to c$ is carried out, too.


\section{Static shear viscosity}

\begin{figure}[b]
  \vspace{.5cm}
  \begin{center}
    \epsfig{file=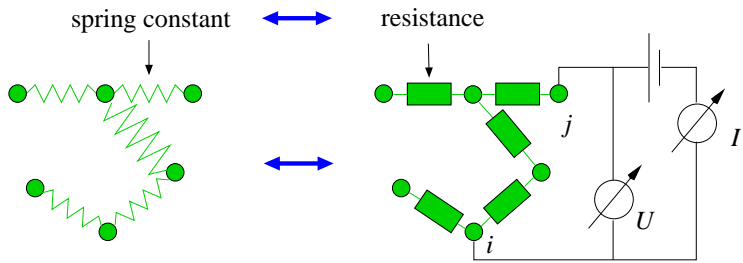, width=0.7\textwidth}
  \end{center}
  \caption{Correspondence between a network of harmonic springs and of
    electrical resistors.}
  \label{mapping}
\end{figure}

According to (\ref{stresstensor}), a time-independent shear rate
$\kappa(t)=\kappa$ induces a time-independent shear stress
$\sigma_{xy}=\rho_0\eta \kappa$
which is determined by the static shear viscosity
\begin{equation}
\label{visco1}
\eta:=\int_0^{\infty} \d t\, \chi(t)=
\frac{\zeta a^{2}}{6N}\,\mathrm{Tr}\left( \frac{1-E_0}{\Gamma}\right)\,.
\end{equation}
As mentioned before, the Rouse model does not include any interactions
between different clusters so that the viscosity allows for a cluster
decomposition. This is apparent in (\ref{visco1}) from the
block-diagonal structure of $\Gamma$. We decompose the graph $\mathcal{ G}$
for a given crosslink realisation into $K$ connected clusters
$\{\mathcal{N}_k\}_{k=1}^K$ with each cluster $\mathcal{ N}_k$
containing $N_k$ 
monomers. The cluster decomposition of the viscosity then reads
\begin{equation}
\eta(\mathcal{ G})=\sum_{k=1}^K \frac{N_k}{N}\eta(\mathcal{ N}_k)\,.
\end{equation}
In order to compute the viscosity of an arbitrary connected cluster
$\mathcal{ N}_k$, we exploit the analogy between a random network of
harmonic springs with spring constants $\lambda_{e}$ and a random
network of resistors of magnitude $1/\lambda_{e}$, see Fig.~\ref{mapping}. 
Since both
networks are governed by linear equations, it is plausible -- and can be
shown exactly \cite{KlRa93} -- that the resistance $\mathcal{ R}(\mathcal{
  N}_k|i,j)$ between any 
two nodes $(i,j)$ in the cluster $\mathcal{ N}_k$ can be expressed in
terms of the pseudo-inverse of the connectivity matrix. This gives rise
to the exact relation \cite{BrLo01}
\begin{equation}\label{relation}
\eta(\mathcal{ N}_k)=\frac{\zeta a^2}{12N_k^2}\sum_{i,j \in \mathcal{ N}_k } 
 \mathcal{ R}(\mathcal{ N}_k|i,j)\,.
\end{equation}
It remains to compute the
average of the viscosity over all realisations of the crosslinks.
For convenience, the average over different crosslink realisations is
performed in two steps: We first average over all 
clusters of a given size $n$ and subsequently average over all cluster
sizes with the appropriate cluster-size distribution $n\tau_n$,
\begin{equation} 
\overline{\eta} = \sum_{n=2}^{\infty} n \tau_n \overline{\eta}_{n}\,.
\end{equation}
Here, $\overline{\eta}_{n}$ is the average of the viscosity over all
clusters of a given size $n$.

\begin{figure}[t]
  \begin{minipage}[t]{0.45\textwidth} 
    \begin{center}
      \epsfig{file=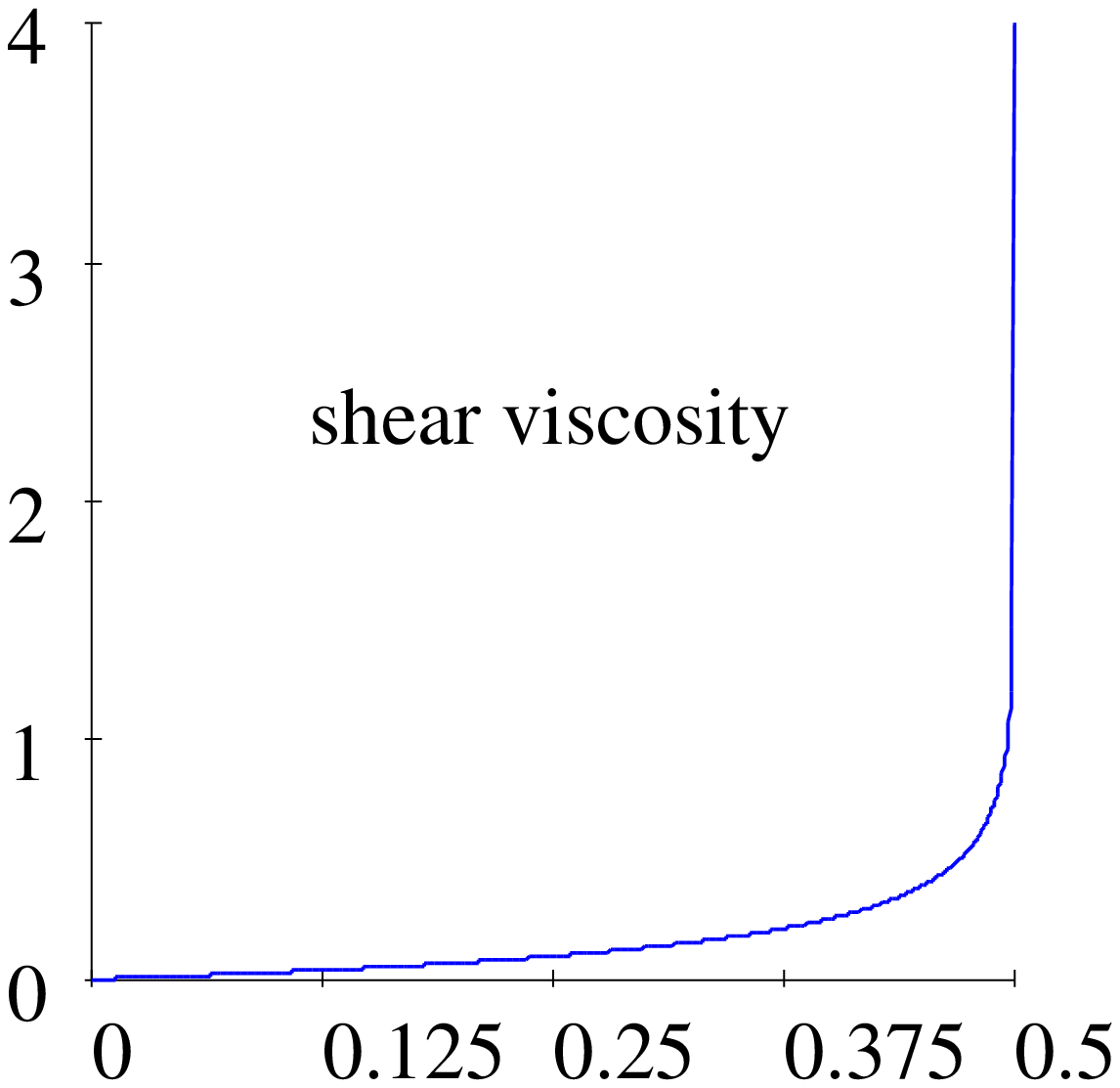, width=\textwidth}
    \end{center}
    \caption{Shear viscosity \protect{(\ref{Eq3.24})} in units of
      $\zeta a^{2}/3$ as a function of $c$ for $P_{1}=1$.}
    \label{viscosityplot}
  \end{minipage} 
  \hfill
  \begin{minipage}[t]{0.45\textwidth}
    \makebox[0cm][l]{
      \hspace*{0.8cm}\epsfig{file=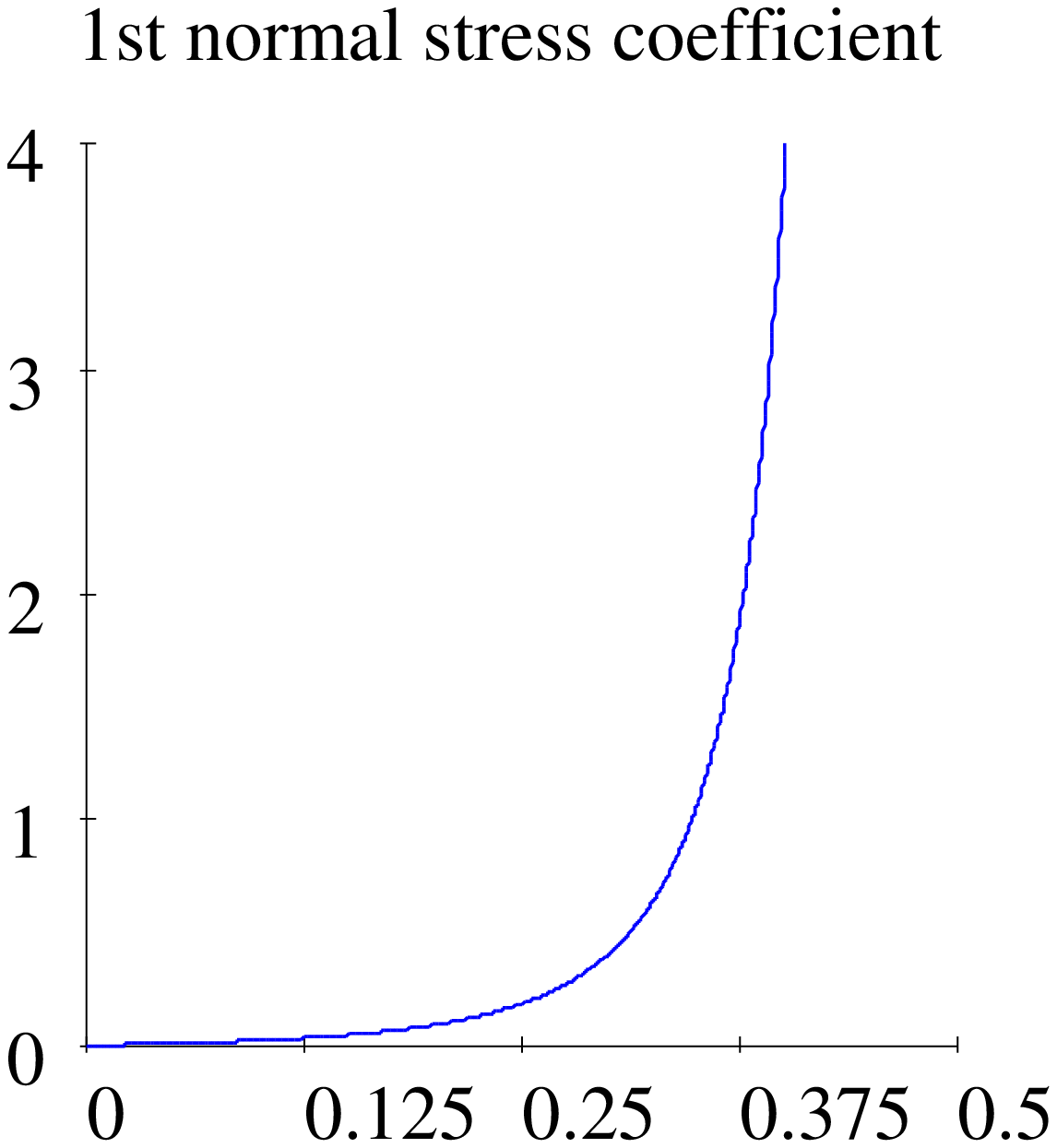, width=\textwidth}
    }
    \caption{First normal stress coefficient (\protect{\ref{psi1erg}})
      in units of $(\zeta a^{2}/3)^{2}$ as a function of $c$ for 
      $P_{1}=P_{2}=1$.}
    \label{normalstressplot}
  \end{minipage}
\end{figure}

First, we consider the ensemble (i) of mean-field random graphs.
For this case, all clusters are almost surely trees \cite{ErRe60}, and
hence the resistance between any two nodes 
$(i,j)$ of a connected cluster is just their weighted chemical distance.
For this reason, and due to the independence of
the distribution of the $\lambda_{e}$, $\overline{\eta}_n$
is known exactly \cite{MeMo70}
\begin{equation}
\overline{\eta}_n=\frac{\zeta a^2}{12} \, P_{1}\, (n-1)!
\sum_{\nu=2}^n \frac{n^{-\nu}\nu(\nu-1)}{(n-\nu)!}
\end{equation}
with $P_{1}$ being defined in (\ref{moments}).
Together with the appropriate cluster-size distribution (\ref{Eq10_3})
we find \cite{BrLo01}
\begin{equation}\label{Eq3.24}  
    \overline{\eta} = 
    \frac{\zeta\, a^2}{24c}\,P_{1}
    \left[ \ln\!\left(\frac{1}{1-2c}\right) - 2c\right]\,.
\end{equation}
Fig.~\ref{viscosityplot} displays $\overline{\eta}$ in
units of $\zeta a^{2}/3$ as a function of $c$ for the special case
$P_{1}=1$. The exact result (\ref{Eq3.24}) is valid for all
$0<c<c_{\mathrm{crit}}=\frac{1}{2}$ and exhibits 
a logarithmic divergence as the percolation transition is approached.

Second, we consider more general percolation ensembles which are only
required  to allow for
a scaling description (\ref{scaling}) close to criticality.
Here we also assume that all spring
constants are fixed, $\lambda_{e}=1$ for $e=1,\ldots,M$. The theory of
random resistor networks \cite{HaLu87,StJaOe99} has established a scaling 
relation
\begin{equation}
\left\langle\frac{1}{N_k}\sum_{i,j\in \mathcal{ N}_k}\mathcal{ R}(\mathcal{ N}_k|i,j)\right\rangle 
\sim (c_{\mathrm{crit}}-c)^{-(2-\eta)\nu-\phi}
\end{equation}
for the resistance which implies \cite{BrLo01} via Eq.\ (\ref{relation})
an analogous 
scaling relation for the viscosity $\langle\eta\rangle_n \sim
n^{\sigma \phi}$ with $\phi$ being the (first) crossover
exponent. Together with the scaling form (\ref{scaling}) of the cluster-size
distribution, we find \cite{BrLo01} for the critical
exponent of the viscosity 
\begin{equation}
  k=\phi-\beta\,.
\end{equation}
For mean-field percolation one has $\phi=\beta=1$, and thus $k=0$, in
accordance with (\ref{Eq3.24}). For 3-dimensional bond
percolation high-precision simulations \cite{GiLo90} for $\phi$ yield
$k\approx 0.71$, in good agreement with recent simulations \cite{VePl01}.


\section{Normal stress coefficients}

The isotropic part of the stress tensor is not significant for the
simple shear flow (\ref{flowfield}). Thus, we concentrate on the first
and second normal stress differences $\sigma_{xx} - \sigma_{yy}$ and 
$\sigma_{yy} - \sigma_{zz}$, respectively. For a time-independent
shear rate $\kappa =\kappa(t)$ it is
customary to define first and second normal stress coefficients by 
\begin{equation}
  \Psi_{1} := \frac{\sigma_{xx} -
    \sigma_{yy}}{\rho_{0}\,\kappa^{2}}\,,\qquad\quad 
  \Psi_{2} := \frac{\sigma_{yy} -
    \sigma_{zz}}{\rho_{0}\,\kappa^{2}}\,.
\end{equation}
One deduces immediately from (\ref{stresstensor}) that $\Psi_{2}=0$, a
characteristic result for Rouse-type models. In contrast, the first
normal stress coefficient $\Psi_{1}$ is non-zero
\begin{equation}
  \label{psi1}
  \Psi_{1} = \frac{1}{2}\,\Bigl(\frac{\zeta a^{2}}{3}\Bigr)^{2}
  \frac{1}{N} \; \mathrm{Tr}\,\Bigl(\frac{1-E_{0}}{\Gamma^{2}}\Bigr)
\end{equation}
and independent of the shear rate $\kappa$. Introducing the (averaged)
density 
\begin{equation}
  \label{dos}
  D(\gamma): =\overline{\frac{1}{N}\,\mathrm{Tr}\,
    \bigl\{(1-E_0)\delta(\gamma -\Gamma)\bigr\}} 
\end{equation}
of non-zero eigenvalues of $\Gamma$, one gets for the
crosslink average of (\ref{psi1})
\begin{equation}
  \overline{\Psi}_{1} = \frac{1}{2}\,\Bigl(\frac{\zeta
    a^{2}}{3}\Bigr)^{2} \int_{0}^{\infty}\!\d\gamma\,
  \frac{D(\gamma)}{\gamma^{2}}\,. 
\end{equation}
The inverse second moment of $D$ was calculated in Eq.\ (38) of
\cite{BrAs00} for mean-field random graphs with the help of a replica
approach. Thus, we infer the exact result
\begin{eqnarray}
  \label{psi1erg}
  \overline{\Psi}_{1} = \frac{1}{2}\,\Bigl(\frac{\zeta
    a^{2}}{3}\Bigr)^{2} &c& \left[  - \frac{8c^{3} - 6c^{2} - 5c +
      1}{30c(1-2c)^{3}}\,P_{1}^{2} - \frac{4c^{2} - 3c
      -1}{24c(1-2c)^{2}} \, P_{2} \right.\nonumber\\
    & &\phantom{[}
    +\left. \frac{5P_{2}-4P_{1}^{2}}{240c^{2}}\,\ln(1-2c)\right]\,,   
\end{eqnarray}
which is valid for all $0<c<c_{\mathrm{crit}}=\frac{1}{2}$. The
moments $P_{n}$ were defined in (\ref{moments}). The result
(\ref{psi1erg}) implies the critical divergence 
\begin{equation}
  \overline{\Psi}_{1} \sim \Bigl(\frac{\zeta
    a^{2}}{3}\Bigr)^{2}\, \frac{P_{1}^{2}}{240}\, (c_{\mathrm{crit}} - c)^{-3}
\end{equation}
at the gelation transition, whereas for $c\to 0$ one has
\begin{equation}
  \overline{\Psi}_{1} = \Bigl(\frac{\zeta
    a^{2}}{3}\Bigr)^{2}\, \frac{P_{2}}{8} \,c + \mathcal{ {O}}(c^{2})\,.
\end{equation}
Fig.~\ref{normalstressplot} displays $\overline{\Psi}_{1}$ in units
of $(\zeta a^{2}/3)^{2}$ as a function of $c$  for the special case 
$P_{1}=P_{2}=1$.

\section{Relaxation at finite frequencies}

According to (\ref{eq:stressrelax}) and (\ref{dos}) the averaged 
time-dependent shear relaxation function $\overline{\chi}(t)$ is
related to the density of non-zero eigenvalues of the connectivity
matrix $\Gamma$ by a Laplace transformation
\begin{equation}
\overline{\chi}(t)=\rho_0 \int_0^{\infty}\d\gamma
D(\gamma)\exp\biggl\{-\gamma \,\frac{6t}{\zeta\, a^{2}}\biggr\}\,.
\end{equation}
The eigenvalue density $D(\gamma)$ has been discussed in detail in
\cite{BrAs00}: 
analytically for mean-field random graphs and numerically for 
finite-dimensional 
percolation. Here we just recall its most prominent feature for
mean-field random graphs:
$D(\gamma)$ shows a Lifshits tail for small eigenvalues, as
first suggested by Bray and Rodgers \cite{BrRo88} 
\begin{equation}\label{lifshitz}
D(\gamma) \sim 
\exp\left\{-\left(\frac{\gamma_0 (1-2c)^3}{\gamma}\right)^{1/2}\right\},
\quad \gamma\,\downarrow\,0,
  \quad c<{\textstyle\frac12}\,,
\end{equation}
provided the probability density $p(\lambda)$ of the random spring
constants $\lambda_{e}$ contains no Dirac delta functions and vanishes
sufficiently fast at the origin (see \cite{BrAs00} for details).
In the context of gelation, the small-$\gamma$-behaviour
(\ref{lifshitz}) of $D$ gives rise to a stretched-exponential decay of
the stress relaxation function 
$\overline{\chi}(t)\sim \exp\bigl\{-(t/t^{*})^{1/3}\bigr\}$ for
long times. In other words, it is the soft-mode excitations of the
clusters which determine the stress relaxation at low frequencies.

\section{Outlook}
The understanding of dynamical critical behaviour at the gelation
transition is currently far from being satisfactory. On the one hand,
the experimental data -- if existent -- scatter widely and therefore
do not allow for a serious check of theoretical predictions. On the
other hand, theoretical results, as presented in this Paper within the
framework of a simple Rouse-type model, are likely to be affected by
shortcomings due to the neglect of certain interactions by the
model itself. We only mention the excluded-volume interaction and the
hydrodynamic interaction, both of which are believed to be of
importance for stress relaxation. It is therefore one goal to incorporate
effects of these interactions in future activities.

Second, it is desirable to extend the semi-microscopic approach
advocated here to the dynamics of the gel phase. This will provide
information about stress relaxation in the gel, in particular, the
critical vanishing of the static shear modulus at the
transition. Moreover, being a cluster of macroscopic size, the gel
admits to ask new types of questions. For instance, as initiated in
\cite{BiMo99}, one may examine
the spatial extent of phonon-type excitations in this random network
as a function of the excitation energy. This may reveal a
localization-delocalization transition, a well-known phenomenon from
disordered systems of very different kinds.

\end{document}